\documentclass[amsmath,amssymb,aps,prl,twocolumn,showpacs,superscriptaddress,reprint]{revtex4}

\usepackage{graphicx}
\usepackage{hyperref}
\usepackage[caption=false]{subfig}

\newcommand{\BSONa}{Ba$_3$Si$_6$O$_{12}$N$_{2}$:Eu }
\newcommand{\BSONb}{Ba$_3$Si$_6$O$_{9}$N$_{4}$:Eu }
\newcommand{\BSONaf}{Ba$_3$Si$_6$O$_{12}$N$_{2}$:Eu}
\newcommand{\BSONbf}{Ba$_3$Si$_6$O$_{9}$N$_{4}$:Eu}
\newcommand{\Hosta}{Ba$_3$Si$_6$O$_{12}$N$_{2}$ }
\newcommand{\Hostb}{Ba$_3$Si$_6$O$_{9}$N$_{4}$ }

\newcommand{\Hostbf}{Ba$_3$Si$_6$O$_{9}$N$_{4}$}
\newcommand{\rhomba}{Eu$_1$Ba$_8$Si$_{18}$O$_{36}$N$_{6}$ }

\newcommand{\rhombbf}{Eu$_1$Ba$_8$Si$_{18}$O$_{27}$N$_{12}$}

\begin{document}

\title{Understanding thermal quenching of photoluminescence from first principles.}

\author{S. Ponc\'e}%
\email{samuel.pon@gmail.com}
\affiliation{%
European Theoretical Spectroscopy Facility, Institute of Condensed Matter and Nanosciences, Universit\'e catholique de Louvain, Chemin des \'etoiles 8, bte L07.03.01, B-1348 Louvain-la-neuve, Belgium.
}%
\author{Y. Jia}
\affiliation{%
European Theoretical Spectroscopy Facility, Institute of Condensed Matter and Nanosciences, Universit\'e catholique de Louvain, Chemin des \'etoiles 8, bte L07.03.01, B-1348 Louvain-la-neuve, Belgium.
}%
\author{M. Giantomassi}
\affiliation{%
European Theoretical Spectroscopy Facility, Institute of Condensed Matter and Nanosciences, Universit\'e catholique de Louvain, Chemin des \'etoiles 8, bte L07.03.01, B-1348 Louvain-la-neuve, Belgium.
}%
\author{M. Mikami}
\affiliation{
MCHC R\&D Synergy Center, Inc. 1000, Kamoshida-cho Aoba-ku, Yokohama, 227-8502, Japan
}%
\author{X. Gonze}
\affiliation{%
European Theoretical Spectroscopy Facility, Institute of Condensed Matter and Nanosciences, Universit\'e catholique de Louvain, Chemin des \'etoiles 8, bte L07.03.01, B-1348 Louvain-la-neuve, Belgium.
}%

\date{\today}

\begin{abstract}

Understanding the physical mechanisms behind thermal effects in phosphors is crucial for white light-emitting diodes (WLEDs) applications,
as thermal quenching of their photoluminescence might render them useless.
The two chemically close Eu-doped \Hosta and \Hostb crystals are typical phosphors studied for WLEDs. The first one sustains efficient light emission at 100$^{\circ}$C while the second one emits very little light at that temperature. Herein, we analyze from first principles their electronic structure and atomic geometry, before and after absorption/emission of light.
Our results, in which the Eu-5d levels are obtained inside the band gap thanks to the removal of an electron from the 4f$^7$ shell, attributes the above-mentioned experimental difference to an auto-ionization model of the thermal quenching, based on the energy difference between Eu$_{\text{5d}}$ and the conduction band minimum.
For both Eu-doped phosphors, we identify the luminescent center, and we show that the atomic relaxation in their excited state is of crucial importance for a realistic description of the emission characteristics.

\end{abstract}


\maketitle

\begin{figure}
 \centering
    \includegraphics[width=0.49\textwidth]{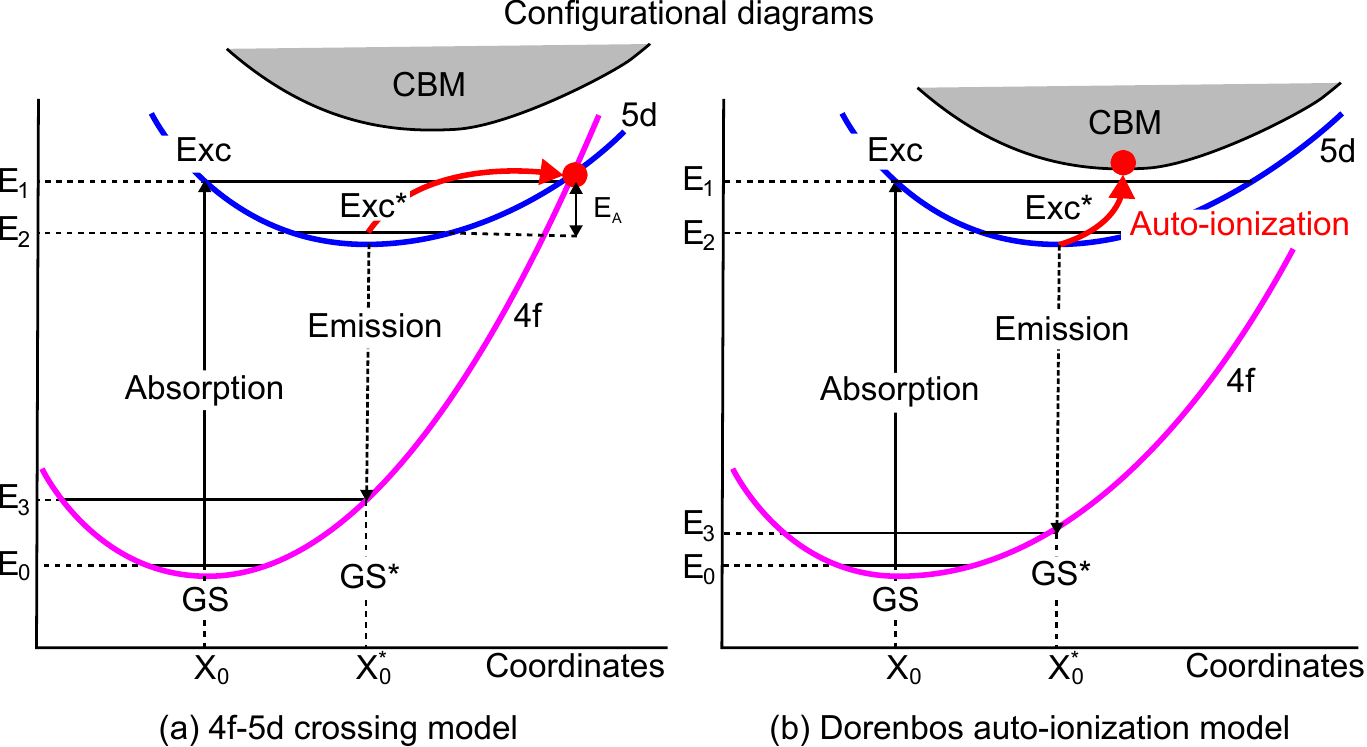}
  \caption[]{\label{fig:Config_Doren} Comparison between the 4f-5d crossing and the Dorenbos auto-ionization model to explain the thermal quenching. }
\end{figure}

White light-emitting diodes (WLEDs) are seen as the most promising light source to replace incandescent and ultraviolet (UV) lamps.
The phosphor-converted WLEDs rely on multiple layers of rare-earth (RE) doped photoluminescent materials called ``phosphors" to down-convert (Stokes shift) the monochromatic UV or blue light 
from an electroluminescent diode into a broad-spectrum white light. The photoluminescent properties in RE-doped phosphors are most often based on radiative 5d-4f transition of the dopant. 
Phenomenological models of the luminescence and Stokes shift of such phosphors are based on configurational diagrams~\cite{Henry1968,Barnett1970} (see Fig.~\ref{fig:Config_Doren}). In the case of RE-doped phosphors, one electron is excited from the RE-4f band into the excited RE-5d band by a UV/blue photon coming from the diode (absorption). Such excited RE-5d electron modifies the forces inside the material, changing the average atomic positions from $X_0$ to $X_0^*$. After dissipation of the energy as phonons, the excited electron spontaneously emits light and the crystal returns to its original ground-state geometry after yet another relaxation. 

Unfortunately for the WLEDs industry, other non-radiative mechanisms allow the RE-5d electron to relax to the ground state. For example, the electron can lose entirely its energy as heat by emitting phonons or transfer its energy to impurities or defects called ``killer-center" via, for example, photo-ionization~\cite{Khan2015}. A phosphor can have high photoluminescence efficiency at low temperature but little at working temperature. The existence of such thermal quenching seem to be sensitive to minute details of the host material and RE dopants.

A striking example is given by two chemically close oxynitride materials: \BSONa and \BSONb (named generically BSON compounds). The two hosts materials have very similar electronic and structural properties~\cite{Bertrand2013} but when doped with Eu, they exhibit different luminescence properties and temperature behavior. The green phosphor \BSONaf~\cite{Uheda2007,Mikami2009,Zhang2010,Braun2010,Tang2011,Porob2011,Song2011,Kang2012,Chen2013,Chen2013a,Li2014} has an intensity that stays almost constant with temperature up to its working temperature, while \BSONb is a bluish-green phosphor~\cite{Stadler2006,Mikami2009,Song2013,Li2013,Kim2013} that exhibits an unfavorable decrease of the luminescence intensity with temperature.

The experimental absorption and emission spectra as well as their temperature dependence have been measured experimentally~\cite{Uheda2007} and lead to a full width at half maximum (FWHM) of the two emission curves at 4~K of 1500~cm$^{-1}$ and 1300~cm$^{-1}$ for Ba$_3$Si$_6$O$_{12}$N$_{2}$:Eu and Ba$_3$Si$_6$O$_{9}$N$_{4}$:Eu, respectively. These two FWHM are considered rather narrow and therefore it is expected that the luminescence in both compounds only comes from one luminescent center (one non-equivalent crystallographic position), despite the existence of
several Ba inequivalent sites (two in \BSONa and three in \BSONb), where substitution by Eu might occur.

In this paper, we clarify the physical origin of the thermal quenching in these RE-doped oxynitrides, by complementing the available experimental data with first-principle results. Additionally, we  confirm the uniqueness of non-equivalent luminescent sites that are present in both doped materials, and identify them. We believe our results are general and concern other RE-doped phosphors. 
Due to the size of the cells needed to represent the doped materials, the Bethe-Salpeter equation (BSE)~\cite{Onida2002}, presently the best technique to predict optical properties in solids, could not be straightforwardly used. 
We rely instead on a simpler mean-field approach, in which a fixed 4f hole is introduced, either frozen in the core by means of a pseudopotential, or artificially forced inside the band gap so that the Eu$_{5d}$ electron is located inside the band gap. This approach allows us to draw, for the first time, a global and coherent picture of the absorption/emission, the geometry relaxation and the temperature effect on the luminescence. 

\paragraph{Models for the thermal quenching.}
The 5d-4f crossing decay model (Fig.~\ref{fig:Config_Doren}a) predicts that large atomic geometry changes lead to the crossing of the Eu$_{5d}$ and Eu$_{4f}$ configurational energy curves. If the temperature is large enough, the excited electron of the 5d band overcome the activation energy barrier $E_A$ and decays non-radiatively to the ground state. 
Based on this model, Blasse and Grabmaier~\cite{Blasse1994} made predictions about the thermal quenching: stronger non-radiative decays occur when the zero phonon line energy difference $E_2-E_0$ is small and the parabola offset as well as the vibrational frequencies are important.  As pointed out by Mikami and co-worker~\cite{Mikami2009,Mikami2011,Mikami2013}, the Ba$_3$Si$_6$\-O$_{12}$N$_2$:Eu compound has a smaller zero phonon line, equivalent parabola offset and larger phonon frequencies than Ba$_3$Si$_6$O$_{9}$\-N$_4$\-:Eu, in contradiction to the observed thermal quenching.  

In contrast, Dorenbos~\cite{Dorenbos2005c} proposed an auto-ionization model, see Fig.~\ref{fig:Config_Doren}b, arguing that the Eu$_{5d}$ comes close to the conduction band minimum (CBM) from the host material well before the 5d-4f crossing point. We will show later that indeed this situation can be inferred from first-principles calculations. At working temperature, the Eu$_{5d}$ electron is transferred to the conduction band, becomes delocalized and mobile, and undergoes energy dissipation through other non-radiative mechanisms (trapping, killer centers such as defects, lattice vibration ...). 
Based on this model, the smaller the Eu$_{\text{5d}}$-CBM gap, the larger is the thermal quenching. 
Dorenbos' empirical formula~\cite{Dorenbos2005c} $\text{CBM}-\text{Eu}_{5d} =\frac{T_{0.5}}{680} eV$ links the size of this gap to the quenching temperature T$_{0.5}$, at which the emission intensity has dropped by 50\% from its low temperature value. From experimental thermal quenching data of the emission spectrum, Mikami~\textit{et al.}~\cite{Mikami2011} could estimate the Eu$_{\text{5d}}$ to CBM gap to be approximatively 0.6 and 0.2~eV for Ba$_3$Si$_6$O$_{12}$N$_2$:Eu and Ba$_3$Si$_6$O$_{9}$N$_4$:Eu, respectively.

\begin{figure}
  \centering
    \includegraphics[width=0.5\textwidth]{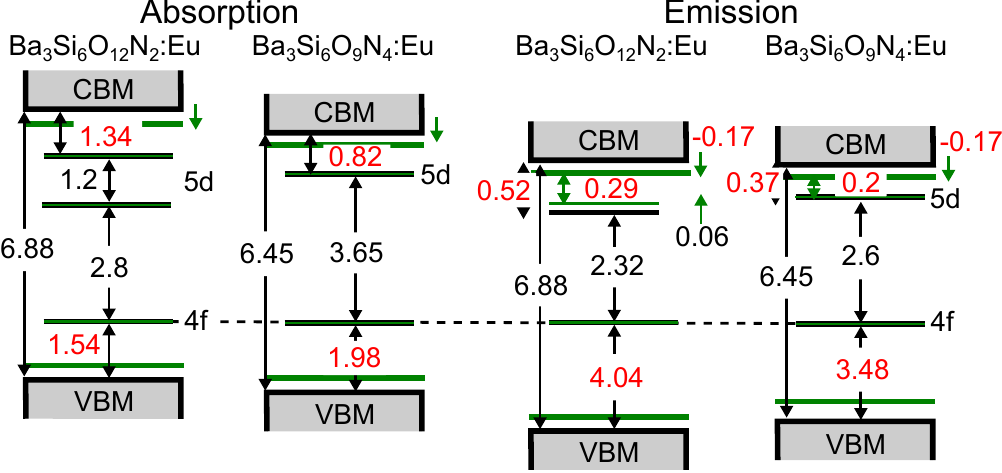}
  \caption[]{\label{fig:Dorenbos_model} Absorption and emission levels (eV) of the two BSON luminescent centers from experimental data or $G_0W_0$ calculations (in black) and present calculations (in red). The horizontal band positions at 0~K are shown in black, and at  300~K are shown in green (sometimes superimposed when there is no thermal effect).}%
\end{figure}

To validate the Dorenbos' model, we have gathered experimental data and employed accurate many-body $G_0W_0$ \textit{ab-initio} bandgaps of 6.88~eV and 6.45~eV for the two host materials~\cite{Bertrand2013}. This allowed us to build a global picture of the level positions for the absorption and emission in the two compounds, see Fig.~\ref{fig:Dorenbos_model}.
For the absorption, the 4f-5d gaps of 310~nm (4~eV) and 440~nm (2.8~eV) in Ba$_3$Si$_6$\-O$_{12}\-$N$_2$:Eu come from the peak and shoulder positions of the experimental absorption spectrum~\cite{Uheda2007}. 
The equivalent 4f-5d gap in Ba$_3$Si$_6$\-O$_{9}\-$N$_4$:Eu is located at 340~nm (3.65~eV). Similar consideration leads us to 5d-4f gaps of 535~nm (2.32~eV) and 480~nm (2.6~eV) for Ba$_3$Si$_6$\-O$_{12}\-$N$_2$:Eu and Ba$_3$Si$_6$\-O$_{9}\-$N$_4$:Eu, respectively. There is an additional blue shift of 0.06~eV at 300~K in the emission curve in the case of Ba$_3$Si$_6$\-O$_{12}\-$N$_2$:Eu~\cite{Uheda2007}. Fig.~\ref{fig:Dorenbos_model} is then confirmed and completed (in red) by our first-principles results, as explained below. 

However, let us emphasize the meaning of the energy changes that are gathered in Fig.~\ref{fig:Dorenbos_model}. At variance with standard density-functional theory (DFT) band structures~\cite{Martin2004} or even $GW$ quasi-particle band structures~\cite{Aulbur1999}, the levels inside the gap are not associated to the removal or addition of an electron but instead to neutral excitations.
The Eu$_{\text{4f}}$ level is occupied by seven electrons in the ground state (half-filled shell). The absorption of a photon leaves a hole in this Eu$_{\text{4f}}$ shell and the electron that is transferred in the 5d lower levels or in the CBM feels the attractive potential of the hole, which 
is absent in traditional band structure methods. On the other hand, the band edges keep their usual significance: they weakly feel the localized 4f hole, since the corresponding electron (or hole) band edge states are delocalized. 

\paragraph{The ground state.}
The crystal structure of the two hosts, where the Ba atom has two inequivalent positions in \Hosta and three in Ba$_3$Si$_6$O$_{9}$N$_{4}$, 
is presented in Ref.~\cite{Bertrand2013}. 
By replacing one Ba atom of the primitive cell by an Eu one, the doping concentration is 33\%, much higher than the 2-10\% experimental doping~\cite{Li2014}. 
Therefore, we relied on doped supercells that are three times larger than the primitive ones, \rhomba and \rhombbf, leading to an Eu doping concentration of 11\%.
More details on the five supercell models are given in the Supplemental Material at [URL will be inserted by publisher]. 
The latter also contains a detailed description of the methodology and numerical parameters of the calculations, whose brief description is given hereafter.

The ground state calculations using the generalized gradient approximation (GGA)~\cite{Perdew1996} +U (U=7.5~eV and J=0.6~eV)~\cite{Liechtenstein1995,Amadon2008} lead to Eu$_{\text{5d}}$ bands that are not localized inside the bandgap of the host material, 
in contradiction to both the 5d-4f crossing decay model and the Dorenbos auto-ionization model.  
For \BSONaf, this had already been noticed in Ref.~\cite{Tang2013}.

\paragraph{Depletion of the 4f shell.}
As already mentioned, the presence of Eu$_{\text{5d}}$ states inside the band gap in Fig.~\ref{fig:Dorenbos_model} is due to
the hole left by the Eu$_{\text{4f}}$ to  Eu$_{\text{5d}}$ transition, creating an attractive center for the Eu$_{\text{5d}}$ electron. 
The Eu$_{\text{5d}}$ electron and the Eu$_{\text{4f}}$ hole are strongly correlated and their description would require a cumbersome treatment at the BSE level.
To make the problem tractable, we use two simpler mean-field approaches that allow us to describe the Coulomb effect of the hole while the Eu$_{\text{5d}}$ state is occupied. 
In the first one, the pseudopotential core-hole approach (see for example Ref.~\cite{Pehlke1993}), a 4f$^6$ shell is frozen in the pseudopotential, instead of the ground state 4f$^7$ shell. 
In the second approach, the Eu$_{\text{4f}}$ hole is treated explicitly in the valence manifold through forced occupation numbers of the Eu$_{\text{4f}}$ orbitals (equal occupation numbers). 
This approach is called  ``constrained DFT'' (CDFT) and has been applied successfully to a wide range of materials~\cite{Dederichs1984,Wu2005a,Wu2006,Kaduk2012}, including rare-earth doped solid-state compounds, by Canning and coworkers, to study the luminescence in Ce-doped~\cite{Canning2011,Chaudhry2011} and Eu-doped~\cite{Chaudhry2014} inorganic scintillators. 
In both approaches, the Coulomb interaction is correctly treated between the Eu core and the explicitly treated Eu$_{\text{5d}}$ electron, as well as 
between the electrons and the other nuclei, allowing to include the most relevant physics in the computation of the forces, and thus yielding reasonable atomic geometries.
We have used the pseudopotential core-hole approach to relax the geometries from first principles. 

 \paragraph{Self-interaction in the excited state.}
However, for the comparison of the 5d electronic level with the CBM, both approaches are quite deceiving, as they include a non-physical self-interaction of the excited electron in the Eu$_{\text{5d}}$ and they poorly describe many-body effects in both Eu$_{\text{5d}}$ and the CBM. The self-interaction overscreens the electronic properties and push the Eu$_{\text{5d}}$ too high in energy, comparatively to the CBM.
 
In order to describe the electronic properties properly, a self-interaction free method has to be used.
We found out that even the simple Kohn-Sham band structure obtained with the depleted Eu$_{\text{4f}}$ shell charge density, but \textit{without occupying the Eu$_{\text{5d}}$ orbitals, nor the conduction band minimum} has no (or little) self-interaction bias between these states.

Moreover, the use of GGA+U is not enough to localize the Eu$_{\text{4f}}$ states inside the bandgap when an electron is removed. An additional atomic potential energy shift of 0.3~Ha for Ba$_3$Si$_6$O$_{12}$N$_2$ and 0.28~Ha for Ba$_3$Si$_6$O$_{9}$N$_4$ has to be applied on the Eu$_{\text{4f}}$ states in order to reproduce the known experimental data shown in Fig.~\ref{fig:Dorenbos_model}. 
Since this additional atomic potential is applied in the same way for the two sets, we can still discuss their relative difference. 
We propose to name our hybrid method ``depleted-shifted 4f approach'' as we remove one electron and apply an additional atomic potential energy to those 4f states. An homogeneous negative background is also added to preserve charge neutrality.

In order to validate it, we have performed tests, for the primitive cell only, based on the BSE and the GW methodology.
In particular, we note that the BSE builds an effective Hamiltonian that takes into account two Feynman diagrams: the direct (attractive Coulomb interaction) diagram and the exchange diagram~\cite{Onida2002}. 
The attractive screened Coulomb interaction obtained from the BSE is accounted for to a large extend in the core-hole and depleted-shifted 4f techniques because the change of the electronic potential resulting  from the hole in the pseudopotential is computed self-consistently within DFT (screened electron-hole interaction) and because all 4f states are strongly localized. 
In contrast, the exchange term $V_{xc}$ within DFT is only a one-point function and therefore does not have the adequate form to reproduce the four-point kernel function of the BSE. 
To evaluate the crudeness of the mean-field approach, we have computed in a restricted space of valence and conduction states, the Bethe-Salpeter exchange term for the Ba$_3$Si$_6$O$_{12}$N$_2$:Eu compounds. We have obtained that the effect of the exchange could be as large as 0.25~eV in high-lying excitonic states, but the lowest transitions were nearly unaffected (0.02~eV). Also, we have performed GW calculations on the doped primitive cell, with the core-hole Eu pseudopotential. The relative changes, between the Eu$_{\text{5d}}$ and the conduction band minimum, compared to the Kohn-Sham eigenenergies,  were much smaller than the GW corrections for the valence to conduction band gap of the host materials.
The latter corrections were on the order of 2.1-2.2 eV for the host materials~\cite{Bertrand2013}, while the GW correction to the Eu$_{\text{5d}}$ to CBM gap were only on the order of 0.2-0.3 eV.

\begin{figure*}
  \centering
    \includegraphics[width=1.0\textwidth]{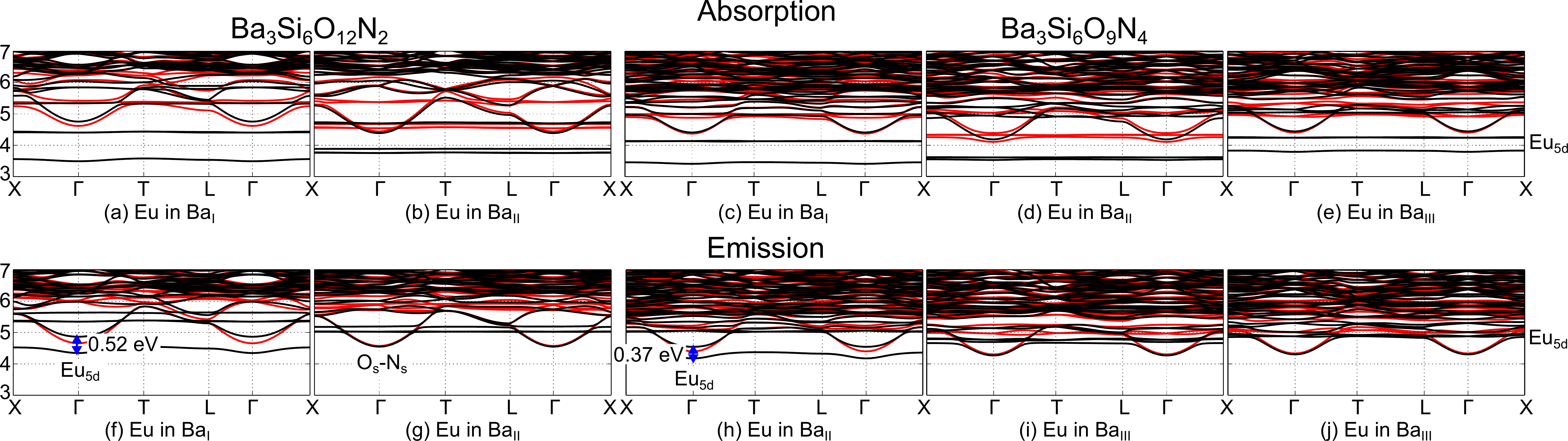}
  \caption[]{\label{fig:Ab_em} Spin-polarized (up in black and down in red) bandstructures corresponding to the absorption (ground-state relaxed atomic positions) and emission (excited-state relaxed positions), in a rhombohedral supercell. One of the nine Ba atoms is replaced by an Eu one at the I, II or III inequivalent Ba positions, leading to an effective Eu doping of 11\%. The zero energy is positioned at the top of the valence band of the host materials and is not shown here. For the emission, the (f) and (h) cases are the luminescent ones whereas the 3 others are not due to the fact that their Eu$_{\text{5d}}$ band is localized above the CBM of the host material. }%
\end{figure*}

\paragraph{The 5d and conduction band levels in the case of neutral excitations.}
The 5d and conduction band levels, obtained with a mean-field 4f hole in the depleted-shifted 4f approach,  were computed for the rhombohedral Eu-doped supercells with different geometries. The results with ground-state atomic positions (absorption) are shown in Figs.~\ref{fig:Ab_em}(a)-(e), while those associated with the emission process, computed at the relaxed atomic positions, are presented in Figs.~\ref{fig:Ab_em}(f)-(j). 
The differences between ground-state and excited geometries are large, and crucial to correctly describe the emission process: in all five cases, the absorption geometry leads to one (or two) Eu$_{\text{5d}}$ level(s) below the CBM (dispersive band), while this only happens in two cases for the relaxed geometry.
The contraction of the O cage around the Eu atom can be extremely important for some bonds, e.g. up to 16\% contraction of the Eu-O bond length with respect to the undoped case. This level of additional contraction is well in line with the change of cation radii associated with the change of oxidation state for the Eu atom that we describe. Indeed, the experimental Eu cation radius goes from 131~pm for Eu$^{2+}$ to 108.7~pm for Eu$^{3+}$~\cite{Shannon1976}.  
From the analysis of the emission in Fig.~\ref{fig:Ab_em}, we conclude that the two Eu$_{\text{I}}$ substitutions in  Ba$_9$Si$_{18}$O$_{36}$N$_{6}$ and Ba$_9$Si$_{18}$O$_{27}$N$_{12}$ are the only luminescent centers due to the position of their Eu$_{\text{5d}}$ band below the CBM, as expected from the small FWHM.
We found out that the two gaps are 0.52~eV and 0.37~eV for Eu$_{\text{I}}$Ba$_8$Si$_{18}$O$_{36}$N$_6$ and Eu$_{\text{I}}$Ba$_8$Si$_{18}$O$_{27}$N$_{12}$, respectively.

For the two luminescent centers, the 4f states (not shown in figure~\ref{fig:Ab_em}), went up by 2.5~eV and 1.5~eV, due to atomic relaxation. 
Although the absolute position of the 4f state depends on the value of the U and the 4f potential shift, the comparison made here is relevant since the only difference between the absorption and emission calculations are the atomic positions.  
To strengthen our analysis, we notice that the lowest conduction bands of the three non-luminescent centers are spin degenerate because the delocalized states are far from the Eu atom and therefore do not react to its magnetic moment. 
We have also made an analysis of the change of 4f state energy with respect to the 5d-CBM gap, for intermediate geometries. We observe that the
5d-CBM gap closes well before the 4f energy comes close to the 5d energy, supporting the Dorenbos model.

\paragraph{The temperature dependence.}
Finally, we  also studied the effect of temperature on the CBM of the two BSON host materials,
 using the static Allen-Heine-Cardona (AHC) theory~\cite{Ponce2014,Ponce2014a,Antonius2014,Marini2015,Ponce2015,Antonius2015}, leading to a similar renormalization of -0.17~eV for both compounds at 300~K. 
 At present, AHC calculations for the doped materials are out of reach computationally. Still, we do not expect large relative differences between the two materials, and in any case much smaller than the -0.17~eV renormalization of the CBM.
 
\paragraph{Conclusion.}
All of the above mentioned results allow us to complete our schematic drawing of Fig.~\ref{fig:Dorenbos_model},
yielding a global picture on which theory and experiment agree.
From Fig.~\ref{fig:Ab_em} we also identify the only  active luminescent center in both \Hosta and \Hostbf.
The three other non-equivalent centers have their Eu$_{\text{5d}}$ states completely above the CBM level of the host material therefore leading to non-radiative emissions. This confirms the experimental results that show narrow and well resolved emission peaks for the two BSON compounds. 
Fig.~\ref{fig:Ab_em} has been obtained through the depleted-shifted 4f scheme, whose validity was checked by comparing it to the Bethe-Salpeter theory in a simplified case.
As seen in Figure \ref{fig:Ab_em}, the relaxation of the Eu doped excited state is of crucial importance as there are major changes in the electronic bandstructure before and after Stokes shifts. 
We confirm the Dorenbos auto-ionization model: the Eu$_{\text{5d}}$-CBM gap is 0.09 eV larger in \BSONa than in \BSONbf. 
This energy difference is enough to show significant thermal quenching behavior differences~\cite{Dorenbos2005c}.   
The techniques that we have used, for the ground-state, the excited state, and finally, for the study of the gap with the conduction band, should be widely applicable for realistic models of phosphors at a much lower computational cost than the much more expensive BSE technique

\paragraph{Acknowledgments.}
We acknowledge discussions with B. Bertrand, D. Waroquiers, and thank J.-M. Beuken for computational help. 
This work has been supported by the Fonds de la Recherche Scientifique (FRS-FNRS Belgium) through a
FRIA fellowship (S.P.) and the PdR Grant No. T.0238.13 - AIXPHO.
Computational ressources have been provided by the supercomputing facilities
of the Universit\'e catholique de Louvain (CISM/UCL)
and the Consortium des Equipements de Calcul Intensif en
F\'ed\'eration Wallonie Bruxelles (CECI) funded by the FRS-FNRS under Grant No. 2.5020.11.

\bibliography{BSON_v18}

\end{document}